\documentclass[aps,prd,showpacs, twocolumn, superscriptaddress, nofootinbib, preprintnumbers]{revtex4}%

\usepackage{amssymb,hyperref, amsmath}
\usepackage[dvips]{color}
\usepackage[dvips]{graphicx}
\usepackage{epsfig}

\begin{document}

\preprint{JLAB-THY-08-875}

\title{Charmonium in lattice QCD and the non-relativistic quark-model}

\author{Jozef J. Dudek}
\affiliation{Jefferson Laboratory MS 12H2, 12000 Jefferson Avenue, Newport News, VA 23606, USA}
\email{dudek@jlab.org}
\affiliation{Department of Physics, Old Dominion University, Norfolk, VA 23529, USA}

\author{Ermal Rrapaj}
\affiliation{Department of Physics, Old Dominion University, Norfolk, VA 23529, USA}

\collaboration{for the Hadron Spectrum Collaboration }

\begin{abstract}
We compare the results of a numerical lattice QCD calculation of the charmonium spectrum with the structure of a general non-relativistic potential model. To achieve this we form the non-relativistic reduction of derivative-based fermion bilinear interpolating fields used in lattice QCD calculations and compute their overlap with $c\bar{c}$ meson states at rest constructed in the non-relativistic quark model, providing a bound-state model interpretation for the lattice data. Essential gluonic components in the bound-states, usually called hybrids, are identified by considering interpolating fields that involve the gluonic field-strength tensor and which have zero overlap onto simple $c\bar{c}$ model states.

\end{abstract}


\maketitle 

\section{Introduction}\label{intro}

New results from CLEO and the $B$-factories have motivated much new interest in the spectrum of charmonium. Several new resonances above open-charm threshold do not appear to fit in the patterns established by the clear sub-threshold states. These lighter states appear to be well described as $c\bar{c}$ states bound in a potential which confines as large distances, but within this framework it has proven difficult to explain many of the new resonances leading to suggestions that we should include physics beyond that allowed in the simplest potential models, specifically that we require a gluonic field carrying non-vacuum quantum numbers (``hybrids'') or higher Fock states such as $cu \bar{u}\bar{c}$ (``multiquarks'' or ``meson-meson molecules''). The clearest signal of physics beyond the quark-model, known as an ``exotic'',  is when a state has external quantum numbers not accessible to a $c\bar{c}$ pair, be that in the $J^{PC}$, such as $1^{-+}$, or in the flavour, such as an isospin 1 state with strong decays to charmed mesons \cite{Abe:2007wga, Mizuk:2008me}.

Much of the phenomenology of the old and new charmonium is performed with quark potential models and their extensions, such as the Coulomb gauge model \cite{Guo:2008yz}, or the flux-tube model \cite{Barnes:1995hc}. These models usually work within a Fock state basis, fixing the number of constituents which have a prescribed set of interactions, usually motivated by, but not formally derived from QCD. The models often predict a mass spectrum and possibly radiative and decay properties which can be compared to experimental data. Assignment of experimental state structure follows if the theory matches the data.

An example of this kind of phenomenology would be the vector ($1^{--}$) channel of charmonium in which established resonances at $3097, 3686, 3770$ MeV and enhancements at $4040, 4160, 4420$ MeV are identified as the model $c\bar{c}$ states with internal quantum numbers $n^{2S+1}L_J = 1^3S_1, 2^3S_1, 1^3D_1, 3^3S_1, 2^3D_1, 4^3S_1$ \cite{Amsler:2008zz}.  The subsequent suggestion of another vector state at a mass $4260$ MeV\cite{Aubert:2005rm, Coan:2006rv, Yuan:2007sj, He:2006kg} is thus supernumerary within this scheme and proposals have been made that one should account for it by considering degrees-of-freedom not present in the quark potential model, such as explicit gluonic quantum numbers, making this a hybrid state, or possibly the quantum admixture of a conventional and a hybrid state\cite{Close:2005iz}. Since the state has external quantum numbers accessible to a $c\bar{c}$ we choose not to label it exotic, with ``crypto-exotic'' implying a hidden exoticness.

Distinguishing between  ``conventional'' and ``crypto-exotic'' states is inherently difficult within lattice QCD calculations; unlike in bound-state models where one must specify upfront the structure of the state, in lattice QCD one simply provides interpolating fields, constructed from quark and gluon fields, of the appropriate external quantum numbers and extracts mass information from QCD correlators. Structure information is embedded in the relative magnitudes of overlap of various interpolating fields on to each state, but it is not always clear how one should interpret this. In this paper we propose a model-dependent scheme based upon a non-relativistic bound-state problem motivated by the physics of quark potential models commonly used to describe heavy quarkonium.

In \cite{Dudek:2007wv}, a set of interpolating fields\footnote{we will also refer to these as ``operators''} built from fermion bilinears with up to two discretised covariant derivatives was used to extract a meson spectrum with charm-mass quarks. The spectral quantities extracted where the masses of the states ($m_N$) and the overlap of each state onto the particular interpolating fields used ($Z_i^{(N)} = \langle 0 | {\cal O}_i(\vec{0},0) | N \rangle $) as expressed in the spectral decomposition of a Euclidean two-point correlator at zero-momentum,
\begin{equation}
 	C_{ij}(t) = 	\sum_{\vec{x}} \langle {\cal O}_i(\vec{x},t) {\cal O}_j(\vec{0},0) \rangle = \sum_N \frac{Z_i^{(N)} Z_j^{(N)*}}{2 m_N} e^{-m_N t}, \nonumber
\end{equation}
where the sum over states labelled by $N$ extends over all meson states with the quantum numbers of the interpolating fields ${\cal O}_{i,j}$. Details of the method used to extract $m,Z$ can be found in \cite{Dudek:2007wv}.

In this paper we compute these overlap factors using a non-relativistic reduction of the continuum limit of the interpolating fields and $c\bar{c}$ states constructed as in a quark potential model. Comparison with the numerical values extracted from the lattice calculation tests the potential model formalism and gives a model-dependent interpretation of the lattice spectrum that agrees reasonably well with the conventional quark-model picture, but goes beyond it by suggesting the influence of gluonic degrees of freedom.

\section{Non-relativistic potential model states}\label{nrstates}

In the non-relativistic potential model we construct states of definite quark and anti-quark number and assume that the only effect of the gluonic field is in providing a static potential which binds the quarks into mesons. A generic unflavoured $q\bar{q}$ meson state can be constructed which is an eigenstate of orbital angular momentum $L$ and total quark spin $S$,
\begin{widetext}
\begin{equation}
	| n^{2S+1}L_J, m_J; \vec{p} \rangle = \sqrt{2 E_{\vec{p}}} \sum_{m_L, m_S} \langle Lm_L; Sm_S| Jm_J\rangle \sum_{r,s} \langle \tfrac{1}{2} r; \tfrac{1}{2} s| Sm_S\rangle     \int \hspace{-2mm} \frac{d^3 \vec{q}}{(2\pi)^3}  \varphi_{nL}(|\vec{q}|) Y^{m_L}_L(\hat{q}) \hspace{2mm} a_r^\dag(\tfrac{1}{2}\vec{p} + \vec{q}) b_s^\dag(\tfrac{1}{2}\vec{p} - \vec{q}) |0\rangle,     \label{modelstate}
\end{equation}
where $a^\dag_r(\vec{p}) / b^\dag_r(\vec{p})$ is the creation operator for a quark/antiquark of momentum $\vec{p}$ and $z$-component of spin $r$. The momentum-space wavefunction $\varphi_{nL}(|\vec{q}|)$, which carries the orbital angular momentum quantum number $L$ and a principal quantum number $n$, is normally determined by solving a Schr\"{o}dinger equation with a potential of phenomenological origin
. Within such models, states of different $L$ can be mixed by non-central interactions such as a tensor force which are usually considered to be relativistic corrections to the dominant central potential, suppressed by powers of $|\vec{q}|/m_q$. 

Although this state is constructed to have non-zero total momentum, it does not transform in a Lorentz-covariant manner under boosts, the model having only Galilean invariance - we will see later that this reduces the usefulness of the model away from states at rest. For the bulk of this paper we will consider only meson states at rest and will leave the momentum-space wavefunctions unspecified.

\end{widetext}

\section{Non-relativistic reduction of interpolating fields}\label{nrred}

A set of operators was presented and used in \cite{Dudek:2007wv}, based upon an extension of the set proposed in \cite{Liao:2002rj}. These operators used a simple discretisation,
\begin{eqnarray}
\overrightarrow{\nabla}_j f(x) &=& \tfrac{1}{2a} \Big( U_j(x) f(x + \hat{j} a) -U_j^\dag(x - \hat{j} a) f(x - \hat{j} a) \Big)  \nonumber \\
 &\to& \overrightarrow{D}_j f(x) + {\cal O}(a^2)\nonumber
\end{eqnarray}
of the covariant derivative and were constructed to transform irreducibly under the group of rotations allowed on a cubic lattice. At zero momentum there are five irreducible representations, $A_1, T_1, T_2, E, A_2$ in which the various continuum spins are distributed \cite{Johnson:1982yq}. The operators used were constructed such that although they transform irreducibly under lattice rotations, they also have a continuum limit in which they overlap with only a single state\footnote{in a few cases there are two continuum overlaps, see the appendix of\protect  \cite{Dudek:2007wv}} of definite $J^{PC}$. These operators then are expected to have an ``unsuppressed'' overlap with one particular $J^{PC}$ along with overlaps with ``lattice artifact'' states, suppressed by powers of the lattice spacing $a$. For example, the operator $\bar{\psi} \overleftrightarrow{\nabla}_i \psi$ has a continuum limit $\bar{\psi} \overleftrightarrow{D}_i \psi$ which overlaps with only $1^{--}$ at zero momentum. At finite lattice spacing this operator transforms as $T_1$ and hence can have overlaps with $3^{--}, 4^{--}\ldots$ suppressed by at least one power of $a$.

We intend to compute overlaps of the type 
\begin{equation}
 	Z = \langle 0 | \bar{\psi}(0) \Gamma \overleftrightarrow{D}_i \overleftrightarrow{D}_j \ldots \psi(0) | n^{2S+1}L_J, m_J; \vec{0} \rangle, \nonumber
\end{equation}
in the limit that the internal momentum of the quarks in the meson is much smaller than the quark mass. We will use the free-field expansion of the quark field operators, $\psi(\vec{x})=$
\begin{equation}
 	 \int \hspace{-2mm}\frac{d^3 \vec{k}}{(2\pi)^3} \sqrt{  \frac{m_q}{\epsilon_{\vec{k}}} } \sum_s \left( a_s(\vec{k}) u_s(\vec{k}) e^{i\vec{k}\cdot \vec{x}}  + b^\dag_s(\vec{k}) v_s(\vec{k}) e^{-i \vec{k}\cdot\vec{x}}     \right)    \nonumber 
\end{equation}
and, since the quark-model state contains no gluonic field operators, we will neglect the gluonic field part of the covariant derivative\footnote{we will discuss the consequences of this for operators involving the commutation of two covariant derivatives later.}, $D_i = \partial_i + A_i \to \partial_i$. In many-body approaches to QCD, such as the Coulomb-gauge model \cite{Guo:2008yz}, neglecting the transverse part of the gluonic field as a first approximation in this way is natural as the gluons appear with an effective mass such that their involvement in the spectrum is suppressed by a mass gap.

In general the overlaps are found to be proportional to an object
\begin{equation}
 	\int \hspace{-2mm} \frac{d^3 \vec{q}}{(2\pi)^3}  \Big( \varphi_{nL}(|\vec{q}|) Y^{m_L}_L(\hat{q})\Big) \;  (2i \vec{q}_i )\; (2i \vec{q}_j) \ldots \; \bar{v}_s(-\vec{q}) \Gamma u_r(\vec{q}).  \nonumber 
\end{equation}
The spinor contraction is evaluated for the 16 possible gamma matrices using an explicit representation and expanded in powers of $\tfrac{|\vec{q}|}{m_q}$. The integrals over angle can be performed by expressing the components of $\vec{q}$ in terms of $Y^m_L(\hat{q})$'s, and the resulting expression, when combined with the quark spin and spin-orbit Clebsch-Gordans of Eqn \ref{modelstate} gives rise to overlap with only certain quark model states. The overlap is expressed via a single remaining integral over $q \equiv |\vec{q}|$ featuring the unspecified radial momentum-space wavefunction, $\varphi_{nL}(q)$.

\subsection{Quark Smearing}
In \cite{Dudek:2007wv}, smeared quark fields were sometimes used - these have the continuum limit $\exp{\left[ \tfrac{1}{4}\sigma^2 \sum_i D_i D_i \right]} \psi(x)$, where $\sigma$ is called the ``smearing radius''. Considering as a first approximation the covariant derivatives to be conventional derivatives using these fields gives overlaps proportional to
\begin{eqnarray}
 	\int\hspace{-2mm}  \frac{d^3 \vec{q}}{(2\pi)^3}  \Big( \varphi_{nL}(|\vec{q}|) Y^{m_L}_L(\hat{q})\Big)\; e^{- \sigma^2 |\vec{q}|^2 / 2} \nonumber \\ \times  (2i \vec{q}_i )\; (2i \vec{q}_j) \ldots \; \bar{v}_s(-\vec{q}) \Gamma u_r(\vec{q}).  \nonumber 
\end{eqnarray}

\begin{figure}
 \includegraphics[width=4.0cm,bb=250 40 680 630]{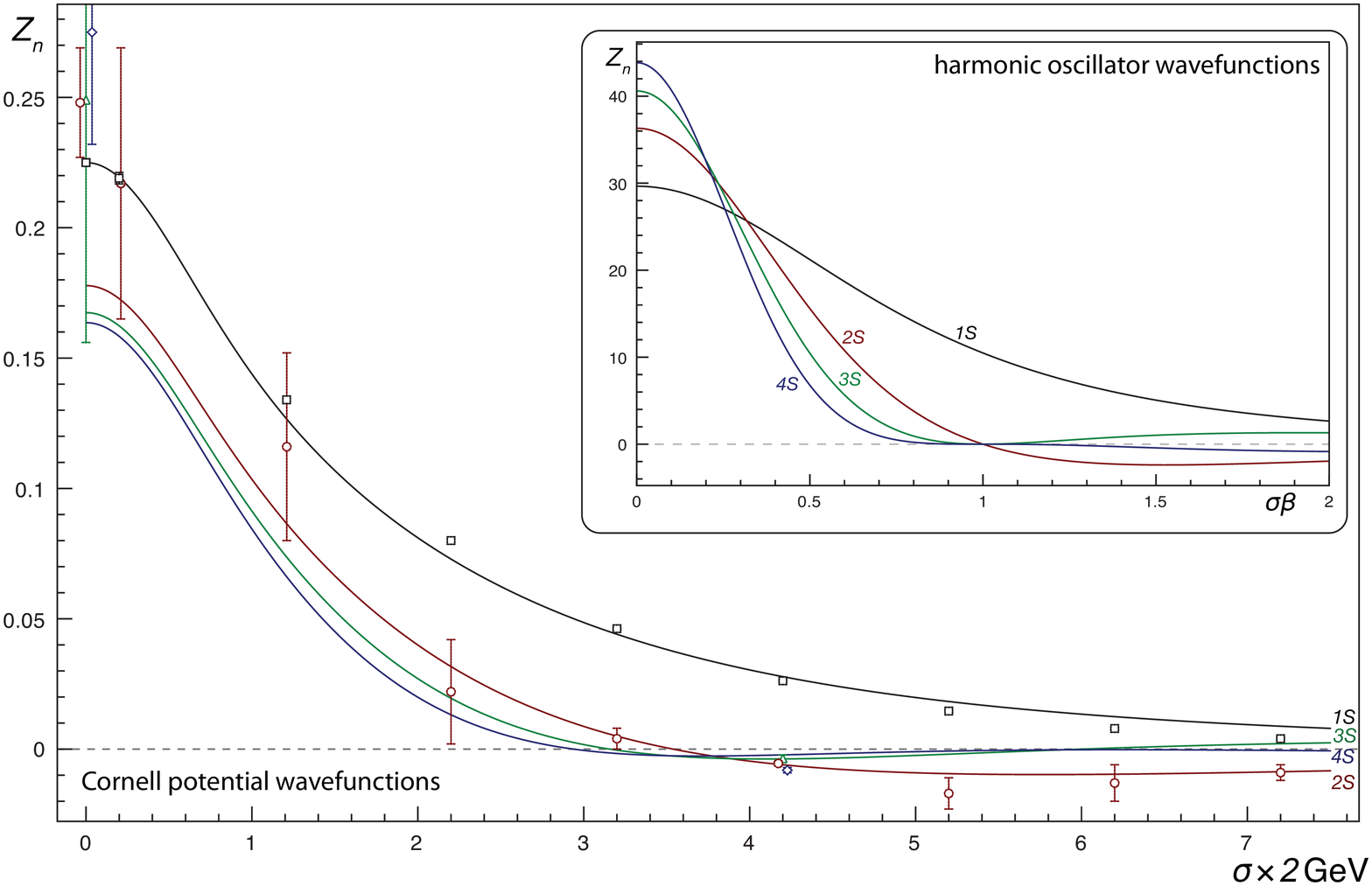}
\caption{Inset: Pseudoscalar overlaps with non-relativistic harmonic oscillator quark model states.
Main Plot: Pseudoscalar overlaps (including the ``relativistic'' correction) with numerical solutions to the Schr\"{o}dinger equation with Cornell potential (solid curves) and lattice QCD simulation data (points).\label{fig:smear}}
\end{figure}

The purpose of smearing the quark fields is to enhance in a correlator the contribution of the ground state as compared to the contributions of excited states. Within the non-relativistic model we can easily demonstrate how this comes about - consider as an example the pseudoscalar states which in this model have a $n^1S_0$ composition. The interpolating field $\bar{\psi} \gamma^5 \psi$ has pseudoscalar quantum numbers\footnote{for the purpose of this argument we are assuming that the lattice is sufficiently fine that states do not resolve the cubic symmetry of the lattice and hence ``lattice artifact'' states can be neglected.} and performing the non-relativistic reduction we find for the overlap with an $n^1S_0$ state, using unsmeared quark fields
\begin{equation}
 	Z_n^{\mathrm{(0)}} = \sqrt{2 M_n}\; 2\sqrt{2\pi} \int \hspace{-2mm}\frac{q^2dq}{(2\pi)^3} \varphi_{nS}(q) \left( 1 + \frac{q^2}{4m_q^2}\right),    \nonumber 
\end{equation}
and using smeared fields
\begin{equation}
 	Z_n^{\mathrm{(\sigma)}} = \sqrt{2 M_n}\; 2\sqrt{2\pi} \int  \hspace{-2mm}\frac{q^2 dq}{(2\pi)^3} e^{-\sigma^2 q^2 / 2} \varphi_{nS}(q) \left( 1 + \frac{q^2}{4m_q^2}\right).    \nonumber 
\end{equation}
The unsmeared expression, if Fourier transformed into position space, is proportional to the value of the spatial wavefunction at the origin with a ``relativistic'' correction proportional to the second derivative of the wavefunction at the origin divided by the quark mass squared.  To get a picture of how smearing might work to enhance the ground state over radially excited states, consider a simple model in which we neglect the relativistic correction and use the bound state wavefunctions of the harmonic oscillator potential\footnote{We'll parameterise this such that the ground state $\sim \exp -\tfrac{q^2}{2\beta^2}$.}. In such a model we can perform the above integrals exactly to find
\begin{equation}
 	Z_n^{\mathrm{(\sigma)}} \propto \left(1-\beta^2\sigma^2\right)^{n-1},   \nonumber 
\end{equation}
so that if we smear with $\sigma = \beta^{-1}$ we have overlap with only the ground state and with none of the excited spectrum. The dependence upon $\sigma \beta$ for the lowest four states is shown in the inset of Figure \ref{fig:smear}. For charmonium a more realistic basis of wavefunctions are the solutions to the Schr\"{o}dinger equation with the Cornell potential, $-\tfrac{4}{3} \tfrac{\alpha}{r} + b r$ - numerically solving this Schr\"{o}dinger equation with the parameters given in \cite{Barnes:2005pb} we find the overlaps shown by the curves in figure \ref{fig:smear}. It is clear that there is still a region of $\sigma$ in which the excited state contributions are much suppressed relative to the ground state.  Also shown are overlaps extracted from a lattice QCD calculation (with the same lattice parameters as those used in \cite{Dudek:2007wv}) for various smearing radii\footnote{for simplicity we normalise such that the unsmeared ground state overlap is equal in the lattice and model calculations.}, where we see that there is generally reasonable agreement between the lattice smearing dependence and the model smearing dependence.

\section{Quark model interpretation of charmonium lattice data}
We have computed the overlap of the continuum, non-relativistic limit of the interpolating fields used in \cite{Dudek:2007wv} with non-relativistic quark model states - the full results are tabulated in Tables \ref{tab:local}, \ref{tab:nabla}, \ref{tab:D}. 
These results should be supplemented with the fact that all of the $\mathbb{B}$-type operators used in \cite{Dudek:2007wv} have zero overlap with any quark model state, which follows from $\mathbb{B} \sim [D,D]  \xrightarrow{A \to 0} 0$. A parenthetical remark is that the operators with $\rho$ and $\rho_2$ in their names can be added/subtracted to produce operators with leading/subleading non-relativistic behaviour. This just corresponds to the usual notion of large upper/small lower components of the non-relativistic limit of a Dirac spinor.

Throughout this paper we use the following notation for the polarisation vectors of a spin-1 particle having momentum $\vec{p}$ and $z$-component of spin $r$: $\in^\mu\hspace{-2mm}(\vec{p},r)$. The polarisation tensors for spin-2 and spin-3 particles are similarly denoted.

\begin{table*}
 \begin{tabular}{c c| c c| c cc}
$\substack{\mathrm{operator}\\ \mathrm{name}}$	& $\substack{\mathrm{continuum}\\ \mathrm{limit}}$	& $\substack{\mathrm{allowed}\\J^{PC}}$ & $\substack{\mathrm{kinematic} \\ \mathrm{factor}}$			& $\substack{\mathrm{quark\, model}\\ \mathrm{state}}$	 & $f(q)$	 & $\substack{\mathrm{origin}\\ \mathrm{behaviour}}$\\
\hline \hline
$1$				& $\bar{\psi} \psi$			& $0^{++}$		& $Z$				& $^3P_0$			& $\sqrt{2} \tfrac{q}{m}$			& $R'_P(0)$ \\
$\gamma^0$		& $\bar{\psi} \gamma^0 \psi$ & $0^{+-}$		& $Z$				& exotic				& 0 & 0 \\
$\gamma^5$		& $\bar{\psi} \gamma^5 \psi$  & $0^{-+}$	& $Z$				& $^1S_0$			& $2\sqrt{2}\left( 1+ \tfrac{q^2}{4m_q^2}\right)$ & $R_S(0)$ \\
$\gamma^0 \gamma^5$		& $\bar{\psi} \gamma^0 \gamma^5 \psi$  & $0^{-+}$	& $Z$				& $^1S_0$			& $2\sqrt{2}\left( 1- \tfrac{q^2}{4m_q^2}\right)$ & $R_S(0)$ \\
$\gamma^i$		& $\bar{\psi} \gamma^i \psi$  & $1^{--}$	& $Z\in^i$				& $^3S_1$			& $2\sqrt{2}\left( 1+ \tfrac{q^2}{12m_q^2}\right)$ & $R_S(0)$ \\
&&&&$^3D_1$ & $\tfrac{2}{3} \tfrac{q^2}{m_q^2}$ & $R_D''(0)$\\
$\gamma^0 \gamma^i$		& $\bar{\psi} \gamma^0 \gamma^i \psi$  & $1^{--}$	& $Z\in^i$				& $^3S_1$			& $2\sqrt{2}\left( 1- \tfrac{q^2}{12m_q^2}\right)$ & $R_S(0)$ \\
&&&&$^3D_1$ & $\tfrac{2}{3} \tfrac{q^2}{m_q^2}$ & $R_D''(0)$\\
$\gamma^i \gamma^5$		& $\bar{\psi} \gamma^i \gamma^5 \psi$  & $1^{++}$	& $Z\in^i$				& $^3P_1$			& $\tfrac{2}{\sqrt{3}} \tfrac{q}{m_q}$ & $R_P'(0)$ \\
$\gamma^i \gamma^j$		& $\epsilon^{ijk} \bar{\psi} \gamma^j \gamma^k \psi$  & $1^{+-}$	& $Z\in^i$				& $^1P_1$			& $\tfrac{2\sqrt{2}}{\sqrt{3}} \tfrac{q}{m_q}$ & $R_P'(0)$ 
 \end{tabular} 
\caption{Local operators. Indicated are the $J^{PC}$ allowed at rest by Lorentz symmetry. The quark-model overlaps are given by $Z = \sqrt{2M} \sqrt{\pi} \int \tfrac{q^2 dq}{(2\pi)^3} \varphi(q) f(q)$ for unsmeared operators and $Z^{(\sigma)} = \sqrt{2M} \sqrt{\pi} \int \tfrac{q^2 dq}{(2\pi)^3} e^{-\sigma^2 q^2 / 2} \varphi(q) f(q)$    for smeared operators. The final column indicates the leading behaviour in terms of derivatives of the spatial wavefunction at the origin. \label{tab:local}}
\end{table*}

\begin{table*}
 \begin{tabular}{c c| c c| c cc}
$\substack{\mathrm{operator}\\ \mathrm{name}}$	& $\substack{\mathrm{continuum}\\ \mathrm{limit}}$	& $\substack{\mathrm{allowed}\\J^{PC}}$ & $\substack{\mathrm{kinematic} \\ \mathrm{factor}}$			& $\substack{\mathrm{quark\, model}\\ \mathrm{state}}$	 & $f(q)$	 & $\substack{\mathrm{origin}\\ \mathrm{behaviour}}$\\
\hline \hline
$a_0 \times \nabla$ 	& $\bar{\psi} \partial^i \psi$	& $1^{--}$ &	$M Z\in^i$	&	$^3S_1$	&	$\tfrac{2\sqrt{2}}{3 M} \tfrac{q^2}{m_q}$ & $R_S''(0)$ \\	
&&&& $^3D_1$ & $\tfrac{4}{3 M} \tfrac{q^2}{m_q}$ & $R_D''(0)$ \\	
$a_{0(2)}\times \nabla$ & $\bar{\psi} \gamma^0 \partial^i \psi$ & $1^{-+}$ & $M Z \in^i$ & exotic & 0 & 0\\
$\pi \times \nabla$ & $\bar{\psi} \gamma^5 \partial^i \psi$ & $1^{+-}$ & $M Z \in^i$ & $^1P_1$ & $\tfrac{4\sqrt{2}}{\sqrt{3}M}q\left( 1+ \tfrac{q^2}{4m_q^2}\right)$  & $R_P'(0)$\\
$\pi_{(2)} \times \nabla$ & $\bar{\psi} \gamma^0 \gamma^5 \partial^i \psi$ & $1^{+-}$ & $M Z \in^i$ & $^1P_1$ & $\tfrac{4\sqrt{2}}{\sqrt{3}M}q \left( 1- \tfrac{q^2}{4m_q^2}\right)$  & $R_P'(0)$\\
$\rho \times \nabla$ & $\bar{\psi} \gamma^i \partial^j \psi$ & $0^{++}$ & $M Z \delta^{ij}$ & $^3P_0$ & $\tfrac{4\sqrt{2}}{3 M}q \left( 1 - \tfrac{q^2}{4m_q^2}\right)$  & $R_P'(0)$\\
&&$1^{++}$& $MZ\epsilon^{ijk} \in^k$ & $^3P_1$ & $\tfrac{4}{\sqrt{3} M}q \left( 1 + \tfrac{q^2}{4m_q^2}\right)$  & $R_P'(0)$\\
&&$2^{++}$& $MZ \in^{ij}$ & $^3P_2$ & $\tfrac{4\sqrt{2}}{\sqrt{3} M}q \left( 1 + \tfrac{q^2}{20 m_q^2}\right)$  & $R_P'(0)$\\
&&&&$^3F_2$ & $\tfrac{4}{5M} \tfrac{q^3}{m_q^2}$  & $R_F'''(0)$ \\

$\rho_{(2)} \times \nabla$ & $\bar{\psi} \gamma^0 \gamma^i \partial^j \psi$ & $0^{++}$ & $M Z \delta^{ij}$ & $^3P_0$ & $\tfrac{4\sqrt{2}}{3 M}q \left( 1 + \tfrac{q^2}{4m_q^2}\right)$  & $R_P'(0)$\\
&&$1^{++}$& $MZ\epsilon^{ijk} \in^k$ & $^3P_1$ & $\tfrac{4}{\sqrt{3} M}q \left( 1 - \tfrac{q^2}{4m_q^2}\right)$  & $R_P'(0)$\\
&&$2^{++}$& $MZ \in^{ij}$ & $^3P_2$ & $\tfrac{4\sqrt{2}}{\sqrt{3} M}q \left( 1 - \tfrac{q^2}{20 m_q^2}\right)$  & $R_P'(0)$\\
&&&&$^3F_2$ & $\tfrac{4}{5M} \tfrac{q^3}{m_q^2}$  & $R_F'''(0)$ \\

$a_1 \times \nabla$ & $\bar{\psi} \gamma^5 \gamma^i \partial^j \psi$ & $0^{--}$ & $M Z \delta^{ij}$ & exotic & 0 & 0 \\
& & $1^{--}$ & $M Z \epsilon^{ijk} \in^k$ & $^3S_1$ & $\tfrac{2\sqrt{2}}{ 3 M} \tfrac{q^2}{m_q}$ & $R_S''(0)$ \\
& & & & $^3D_1$ & $\tfrac{2}{ 3 M} \tfrac{q^2}{m_q}$ & $R_D''(0)$ \\
& & $2^{--}$ & $MZ \in^{ij}$ & $^3D_2$ & $\tfrac{2\sqrt{2}}{\sqrt{5}M} \tfrac{q^2}{m_q}$ & $R_D''(0)$\\

$b_1 \times \nabla$ & $ \epsilon^{ikl} \bar{\psi} \gamma^k \gamma^l \partial^j \psi$ & $0^{-+}$ & $M Z \delta^{ij}$ & $^1S_0$ & $\tfrac{4\sqrt{2}}{ 3 M} \tfrac{q^2}{m_q}$ & $R_S''(0)$ \\
& & $1^{-+}$ & $M Z \epsilon^{ijk} \in^k$ & exotic & 0 & 0 \\
& & $2^{-+}$ & $MZ \in^{ij}$ & $^1D_2$ & $\tfrac{8}{\sqrt{15}M} \tfrac{q^2}{m_q}$ & $R_D''(0)$\\

 \end{tabular} 
\caption{As Table \ref{tab:local} for single derivative operators. \label{tab:nabla}}
\end{table*}

\begin{table*}
 \begin{tabular}{c c| c c| c cc}
$\substack{\mathrm{operator}\\ \mathrm{name}}$	& $\substack{\mathrm{continuum}\\ \mathrm{limit}}$	& $\substack{\mathrm{allowed}\\J^{PC}}$ & $\substack{\mathrm{kinematic} \\ \mathrm{factor}}$			& $\substack{\mathrm{quark\, model}\\ \mathrm{state}}$	 & $f(q)$	 & $\substack{\mathrm{origin}\\ \mathrm{behaviour}}$\\
\hline \hline
$a_0 \times \mathbb{D}$ & $|\epsilon^{ijk}| \bar{\psi} \partial^j \partial^k \psi$ & $2^{++}$ &  $ M^2 Z |\epsilon^{ijk}| \in^{jk}$ & $^3P_2$ & $\tfrac{8\sqrt{2}}{5\sqrt{3} M^2} \tfrac{q^3}{m_q}$ & $R_P'''(0)$ \\
&&&& $^3F_2$ &  $\tfrac{8}{5 M^2} \tfrac{q^3}{m_q}$ & $R_F'''(0)$ \\

$a_{0(2)} \times \mathbb{D}$ & $|\epsilon^{ijk}| \bar{\psi} \gamma^0 \partial^j \partial^k \psi$ & $2^{+-}$ &  $ M^2 Z |\epsilon^{ijk}| \in^{jk}$ &exotic & 0 & 0 \\

$\pi \times \mathbb{D}$ & $|\epsilon^{ijk}| \bar{\psi} \gamma^5 \partial^j \partial^k \psi$ & $2^{-+}$ &  $ M^2 Z |\epsilon^{ijk}| \in^{jk}$ & $^1D_2$ & $\tfrac{16}{\sqrt{15}M^2} q^2 \left( 1 + \tfrac{q^2}{4m_q^2}  \right)$ & $R_D''(0)$ \\

$\pi_{(2)} \times \mathbb{D}$ & $|\epsilon^{ijk}| \bar{\psi} \gamma^0 \gamma^5 \partial^j \partial^k \psi$ & $2^{-+}$ &  $ M^2 Z |\epsilon^{ijk}| \in^{jk}$ & $^1D_2$ & $\tfrac{16}{\sqrt{15}M^2} q^2 \left( 1 - \tfrac{q^2}{4m_q^2}  \right)$ & $R_D''(0)$ \\

$\rho \times \mathbb{D}$ & $|\epsilon^{jkl}| \bar{\psi} \gamma^i \partial^k \partial^l \psi$ & $1^{--}$ & $M^2 Z |\epsilon^{ijk}|\in^k$ & $^3S_1$ & $\tfrac{8}{15\sqrt{2}M^2} \tfrac{q^4}{m_q^2}$ & $R_S''''(0)$ \\
& & & & $^3D_1$ & $\tfrac{24}{15M^2}q^2 \left( 1- \tfrac{q^2}{12m_q^2} \right) $ & $R_D''(0)$ \\
& & $2^{--}$ & $M^2 Z |\epsilon^{jkl}| \epsilon^{ikm} \in^{ml }$ & $^3D_2$ & $\tfrac{8\sqrt{2}}{3\sqrt{5} M^2} q^2 \left( 1+ \tfrac{q^2}{4m_q^2}\right)$ & $R_D''(0)$\\
&& $3^{--}$ & $M^2 Z  |\epsilon^{jkl}| \in^{ikl}$ & $^3D_3$ & $\tfrac{16}{\sqrt{15}M^2} q^2 \left( 1+ \tfrac{q^2}{28m_q^2}\right)$ & $R_D''(0)$ \\
&&&& $^3G_3$ & $\tfrac{16}{7\sqrt{5}M^2} \tfrac{q^4}{m_q^2}$ & $R_G''''(0)$\\

$\rho_{(2)} \times \mathbb{D}$ & $|\epsilon^{jkl}| \bar{\psi} \gamma^0 \gamma^i \partial^k \partial^l \psi$ & $1^{--}$ & $M^2 Z |\epsilon^{ijk}|\in^k$ & $^3S_1$ & $\tfrac{8}{15\sqrt{2}M^2} \tfrac{q^4}{m_q^2}$ & $R_S''''(0)$ \\
& & & & $^3D_1$ & $\tfrac{24}{15M^2}q^2 \left( 1+ \tfrac{q^2}{12m_q^2} \right) $ & $R_D''(0)$ \\
& & $2^{--}$ & $M^2 Z |\epsilon^{jkl}| \epsilon^{ikm} \in^{ml }$ & $^3D_2$ & $\tfrac{8\sqrt{2}}{3\sqrt{5} M^2} q^2 \left( 1- \tfrac{q^2}{4m_q^2}\right)$ & $R_D''(0)$\\
&& $3^{--}$ & $M^2 Z  |\epsilon^{jkl}| \in^{ikl}$ & $^3D_3$ & $\tfrac{16}{\sqrt{15}M^2} q^2 \left( 1- \tfrac{q^2}{28m_q^2}\right)$ & $R_D''(0)$ \\
&&&& $^3G_3$ & $\tfrac{16}{7\sqrt{5}M^2} \tfrac{q^4}{m_q^2}$ & $R_G''''(0)$\\

$a_1 \times \mathbb{D}$ & $ |\epsilon^{jkl}|     \bar{\psi} \gamma^5 \gamma^i \partial^k \partial^l \psi $ & $1^{++}$ & $M^2 Z |\epsilon^{ijk}|\in^k$ & $^3P_1$ & $\tfrac{4}{5\sqrt{3}M^2} \tfrac{q^3}{m_q}$ & $R_P'''(0)$ \\
&& $2^{++}$ &  $M^2 Z |\epsilon^{jkl}| \epsilon^{ikm} \in^{ml }$ & $^3P_2$ & $\tfrac{4\sqrt{2}}{5\sqrt{3} M^2} \tfrac{q^3}{m_q} $ & $R_P'''(0)$\\
&&&&$^3F_2$ &  $\tfrac{8}{15 M^2} \tfrac{q^3}{m_q} $ & $R_F'''(0)$\\
&& $3^{++}$ & $M^2 Z  |\epsilon^{jkl}| \in^{ikl}$ & $^3F_3$ & $\tfrac{16}{\sqrt{105}M^2} \tfrac{q^3}{m_q} $ & $R_F'''(0)$ \\

$b_1 \times \mathbb{D}$ & $\epsilon^{imn} |\epsilon^{jkl}|     \bar{\psi} \gamma^m \gamma^n \partial^k \partial^l \psi $ & $1^{+-}$ &  $M^2 Z |\epsilon^{ijk}|\in^k$ & $^1P_1$ & $\tfrac{8\sqrt{2}}{5\sqrt{3} M^2} \tfrac{q^3}{m_q}$ & $R_P'''(0)$ \\
 & & $2^{+-}$ & $M^2 Z |\epsilon^{jkl}| \epsilon^{ikm} \in^{ml }$ & exotic & 0 & 0\\
&& $3^{+-}$ & $M^2 Z  |\epsilon^{jkl}| \in^{ikl}$ & $^1F_3$ & $\tfrac{16}{\sqrt{35}M^2} \tfrac{q^3}{m_q} $ & $R_F'''(0)$ 

 \end{tabular} 
\caption{As Table \ref{tab:local} for two derivative operators.  \label{tab:D}}
\end{table*}

In what follows we will attempt a model-dependent description of parts of the lattice spectrum by comparing the relative size of overlaps extracted from a numerical calculation with the quark model expressions in Tables \ref{tab:local}, \ref{tab:nabla}, \ref{tab:D}. The lattice data used is an extension of that presented in \cite{Dudek:2007wv}, using this time a somewhat larger basis of operators, in particular more operators of high mass-dimension. Doing this allows overlap with discretisation artifact states with a suppression that is smaller in powers of lattice spacing, $a$, so we would expect to see a somewhat more dense spectrum\footnote{In addition, the meson source operator was placed a little further from the temporal Dirichlet wall, reducing the possibility of states propagating forward after reflection off the wall.}. We remind the reader that we expect these lattice results to be dominated by sources of systematic error, notably the use of the quenched approximation, the lack of explicit extrapolation to the continuum and possibly (we shall discuss this later) an overly small spatial volume. Nevertheless we still feel it is a valid exercise to compare the patterns seen in the data with the patterns present in the quark model overlaps. The use of the quenched theory removes the possibility of multiquark states involving light quarks.

\begin{figure}
 \includegraphics[width=6cm,bb=0 0 323 415]{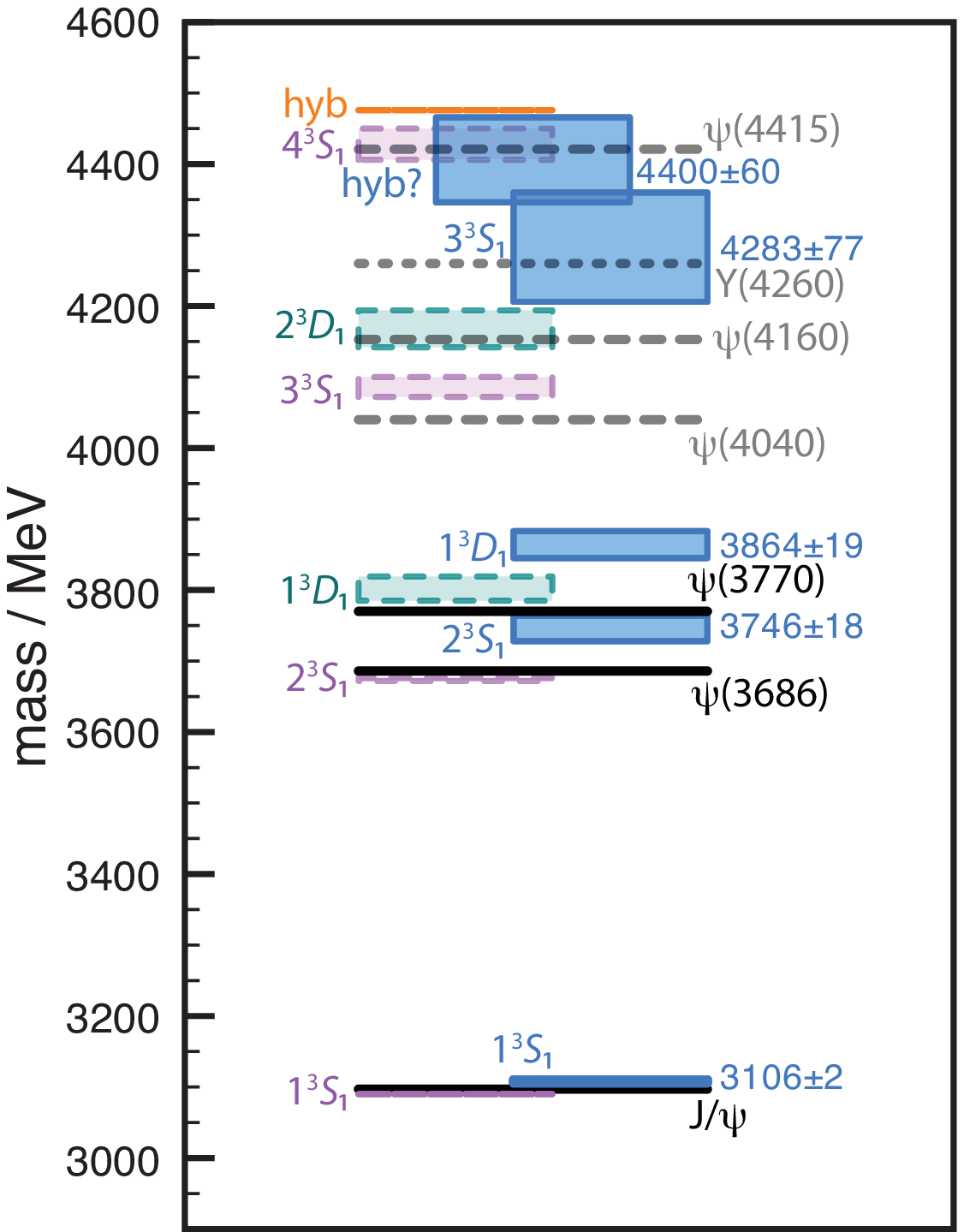}
\caption{$1^{--}$ spectrum. Black lines are experimental states at various levels of confirmation, blue bars are lattice states with quark model assignments as described in the text. Also shown are quark model predictions and the hybrid prediction of the Coulomb gauge model. \label{fig:vector}}
\end{figure}

As an illustrative example, we consider the $T_1^{--}$ channel - at a finite lattice spacing correlators in this symmetry channel will receive contributions from states that in the continuum have $J^{PC}=1^{--}, 3^{--}, 4^{--}\ldots$. With a set of twelve operators we extracted the mass spectrum and overlap values shown in Table \ref{tab:vector} and Figure \ref{fig:vector}. By comparison with the quark model overlap forms we can propose a model-dependent interpretation of the spectrum that bares a strong resemblance to the conventional quark model picture.

The ground state (of mass $3106(2)$ MeV) in this channel has large overlaps with the unsmeared local operators ($\gamma_i, \gamma_0\gamma_i$) indicating that it has a considerable wavefunction at the origin and suggesting a dominantly $^3S_1$ structure in line with what is expected for the $J/\psi$. The first excited state (of mass $3764(18)$ MeV) also has large $Z^{(0)}(\gamma_i, \gamma_0\gamma_i)$ indicating  $^3S_1$, and has a relatively suppressed coupling to the smeared local operators compared to the ground state - from the previous discussion we expect this for a radial excitation and thus propose that this state is dominantly $2^3S_1$. 

The second excited state (of mass $3846(12)$ MeV) has small overlaps onto all the operators used. A likely explanation of this is that we are getting overlap onto a $3^{--}$ state through lattice discretisation effects. In this case the overlaps would be suppressed by powers of the lattice spacing $a$ relative to the overlaps onto the $1^{--}$ states, which are all that these operators have overlap with in the continuum (see the appendix of \cite{Dudek:2007wv}). Within the quark model a $3^{--}(^3D_3)$ state is expected in this mass region and indeed in the $T_2^{--}, A_2^{--}$ channels, where we include operators with continuum $3^{--}$ overlap, there are candidate states whose $Z$ values fit with belonging to a $3^{--}$ state (see \cite{Dudek:2007wv} for details of the spin-assignment procedure which depends only on continuum properties of the operators and not assumptions about models).

The third excited state (of mass $3864(19)$ MeV) has small overlap onto the unsmeared local operators suggesting it is not dominantly a $^3S_1$ state. It does however have a large overlap with the $\rho\times \mathbb{D}_{T1}$ operator which within the quark model has suppressed overlap onto $^3S_1$ relative to $^3D_1$ - this is due to the $\mathbb{D}$ operator transforming like $Y_2^m$. We propose that this state is mostly $1^3D_1$  - its near degeneracy with the $3^{--}$ state reflecting the small spin-orbit interactions in charmonium. It would appear that there is relatively little mixing between the nearby $2S, 1D$ states, corresponding to a small tensor interaction.

The fourth excited state (of mass $4283(77)$ MeV) has large overlap onto the unsmeared local operators suggesting that it may be dominantly $^3S_1$. We suggest that it is the $3^3S_1$ state.

The fifth excited state (of mass $4400(60)$ MeV) has rather small overlaps onto all operators with the notable exception of the smeared $\pi\times \mathbb{B}, \pi_2\times \mathbb{B}$ operators where the overlap is an order of magnitude larger than for the other states so far considered. The $\mathbb{B}$-type operators have the characteristic feature that they are zero under the conditions of the quark model, corresponding as they do to the commutator of two covariant derivatives which vanishes if the gluonic field is neglected. With gluonic field included, the $\mathrm{B}$ operator is proportional to the field strength tensor which in a constituent gluon model would have at least one gluon creation/annihilation operator. Within an extended quark model we would propose that considerable overlap onto $\mathbb{B}$-type operators indicates some hybrid gluonic nature to the state. In the flux-tube model, there is a degenerate set of exotics that includes a $1^{--}$ state at around $4.2$ GeV - within this model the hybrid $1^{--}$ is a quark spin-singlet, which matches with the large overlap onto $\pi \times \mathbb{B}$ which is dominantly spin-singlet. In the Coulomb gauge model \cite{Guo:2008yz} there is a $1^{--}$ hybrid state with a mass near $4.5$ GeV which is also a quark spin-singlet.

\begin{table*}
\begin{tabular}{c|cccccc}
 operator & $0^{\mathrm{th}}[3106(2)]$ &  $1^{\mathrm{st}}[3746(18)]$ &  $2^{\mathrm{nd}}[3846(12)]$  &  $3^{\mathrm{rd}}[3864(19)]$ &  $4^{\mathrm{th}}[4400(60)]$ &  $5^{\mathrm{th}}[4283(77)]$ \\
\hline \hline
$\gamma_i^{(sm)}\;(10^{-3})$ 				&  \textbf{26.8(8) }	& 2.8(6) 	& \textit{0.7(5) }	& \textit{0.8(3)} 	& 3.2(7) 	& 3.5(11) \\
$\gamma_0 \gamma_i^{(sm)}\; (10^{-3})$ 	&  \textbf{26.2(8)}	& 3.3(6)	& \textit{1.1(5)}	& 1.3(4)	& 2.8(8)	& 3.5(11) \\
$a_0\times \nabla_{T1}^{(sm)}\; (10^{-3})$	& 3.53(2)		& 2.4(3)	& 5.3(6)	& 5.7(8)	& \textit{0.2(6)}	& \textit{0.6(3)}	\\
$a_1\times \nabla_{T1}^{(sm)}\; (10^{-3})$	& 6.80(4)		& 4.7(2)	& 5.2(13)	& 7.0(8)	& 3.3(8)	& \textit{1.2(9)} \\
$\rho \times \mathbb{D}_{T1}^{(sm)}\; (10^{-4})$		& 10.4(1)		& \textit{6(6)}	& \textit{7(60)}	& \textbf{290(20) }& \textit{12(29)}	& \textit{14(14)} \\
$\rho_2 \times \mathbb{D}_{T1}^{(sm)}\; (10^{-4})$ 	& 6.1(1)		& \textit{17(6)}	& \textit{20(60)}	& \textbf{290(20) }& \textit{20(20)}	& \textit{7(16) }\\
$\pi \times \mathbb{B}_{T1}^{(sm)}\; (10^{-5})$		& 30.6(5)		& \textit{4(5)}	& \textit{3(7)}	& \textit{3(6)}	& \textbf{130(10) }& \textit{25(25)} \\
$\pi_2 \times \mathbb{B}_{T1}^{(sm)}\; (10^{-5})$		& 28.6(4)		& \textit{8(5)}	& \textit{5(8)}	& \textit{7(6)}	& \textbf{130(10) }& \textit{25(26)} \\
$\gamma_i$								& \textbf{0.163(1)}	& \textbf{0.190(6)} & \textit{0.005(25) }& \textit{0.027(13) }& \textit{0.037(31) }& \textbf{0.202(23) }\\
$\gamma_0 \gamma_i$						& \textbf{0.146(1)}	& \textbf{0.166(6) }& \textit{0.029(20) }& \textit{0.004(14) }& \textit{0.027(24) }& \textbf{0.168(20) }\\
$a_1 \times \nabla_{T1}$					& 0.163(1)	& 0.281(11) & \textit{0.073(47)} & 0.129(27) & \textit{0.08(6) }& 0.391(53) \\
$\pi \times \mathbb{B}_{T1}\;(10^{-3})$				& 8.89(9)		& 8.5(5)	& \textit{1.4(10)	}& \textit{1(1)	}& \textit{8(3)}	&\textit{ 8(3)} \\
$\pi_2 \times \mathbb{B}_{T1}\;(10^{-3})$				& 9.09(9)		& 8.0(5)	& \textit{0.7(10)}	& \textit{0.5(8)	}& \textit{8(3)}	&\textit{ 8(3)}	\\
\hline
assignment	& $1^3S_1$ & $2^3S_1$ & $3^{--}$ ? & $
1^3D_1$ & hybrid ? & $3^3S_1$
\end{tabular} 
\caption{$T_1^{--}$ overlaps from numerical lattice QCD computation. Where an operator has the superscript $(sm)$, the quarkfields have been Gaussian smeared - smearing parameters are given in \protect \cite{Dudek:2007wv}. Numbers in bold are characteristically large, indicating a particular quark model state assignment, while those in italics are both characteristically small and statistically compatible with zero within $3\sigma$. The final row is a model-dependent state assignment described in the text.\label{tab:vector}}
     \end{table*} 

Although the masses do not agree terribly well with experimental candidates it would appear that the spectrum we are observing is in line with the expectation of quark potential models with the intriguing addition of a state which may be hybrid in nature, perhaps being a candidate for the state being claimed in experiment at 4260 MeV. We note here that as well as
 the masses appearing to be systematically high, we also overpredict the vector decay constants (proportional to $Z^{(0)}(\gamma_i)$) for excited states. Although we could not find any finite volume effects in a limited study in \cite{Dudek:2007wv}, we strongly suspect that this is the cause of these effects - if one numerically solves the Schr\"{o}dinger equation with the Cornell potential while forcing boundary conditions corresponding to finite volume and hence ``squeezing'' the wavefunction, one does observe systematically higher masses and larger wavefunctions at the origin high in the spectrum than the solutions in infinite volume. Future lattice work in larger volumes will clear this up.

An example of a channel where the assignment is not so clear is $A_1^{-+}$, whose continuum content is $0^{-+}, 4^{-+},\ldots$. The large overlap, seen in Table \ref{tab:ps}, onto the local operators for all four states considered suggests they all have large $^1S_0$ components. Within the $\bar{c}c$ quark model this is the only possibility for pseudoscalar quantum numbers. The near degeneracy of the second and third excited states is not expected for radial excitations in the Cornell potential. The possibility of a lattice artifact $4^{-+}$ ($^1G_4$ in the quark model, expected around $4.2-4.3$ GeV with the Cornell potential) appears unlikely owing to the large overlaps with local operators. Within the Coulomb-gauge model of \cite{Guo:2008yz} and the flux-tube model, there is a $0^{-+}$ hybrid expected degenerate with a $1^{--}$ hybrid - we proposed such a state at $4400(60)$ in the $T_1^{--}$ analysis above. The second and third states suffer from large statistical fluctuations - modulo this the large overlaps are with the smeared $\rho\times \mathbb{B}$ operator and unsmeared local operators, suggesting hybrid and  $^1S_0$ nature; the $\rho\times \mathbb{B}$ operator is quark spin triplet as is expected of the hybrid state. Within the large fluctuations there is room for some mixing of the conventional and hybrid states.

\begin{table*}
\begin{tabular}{c|cccc}
 operator & $0^{\mathrm{th}}[3027(2)]$ &  $1^{\mathrm{st}}[3714(27)]$ &  $2^{\mathrm{nd}}[4280(60)]$  &  $3^{\mathrm{rd}}[4500(150)]$ \\
\hline \hline
$\gamma_0 \gamma_5^{(sm)}\;(10^{-3})$ 	&  \textbf{24.2(8) }	& 7.0(9) 	& 0.41(14)	& 7.6(8) \\
$\gamma_5^{(sm)}\;(10^{-3})$ 	&  \textbf{26.2(7) }	& 5.5(9) 	& \textit{3.6(13)}  & 6.3(8) \\
$b_1 \times \nabla_{A1}^{(sm)}\;(10^{-3})$ 	& 9.46(5)		& 5.9(5)	& 7.4(21)		& 6.3(8) \\
$\rho \times \mathbb{B}_{A1}^{(sm)}\;(10^{-3})$ 	&  0.9(1)	& \textit{0.27(16)}	& \textbf{2.4(5)}	& \textbf{1.4(5)} \\

$\gamma_0 \gamma_5$ 	&  \textbf{0.162(1) }	&\textbf{ 0.163(13)} 	& \textbf{0.164(63)}	& \textbf{0.179(28)}\\
$\gamma_5$ 	&  \textbf{0.225(1) }	&\textbf{ 0.248(21)} 	& \textbf{0.249(93)}  & \textbf{0.275(42)} \\
$b_1 \times \nabla_{A1}$ 	& 0.281(2)	& 0.471(44)	& 0.52(17)	& 0.64(10) \\
$\rho \times \mathbb{B}_{A1}\;(10^{-2})$ 	&  3.00(3)	& 2.6(2)	& \textit{0.8(17)}	& 3.5(6) \\

\hline
assignment	& $1^1S_0$ & $2^1S_0$ & hyb (+$3^1S_0$) ? & $ 3^1S_0$  (+ hyb) ?
\end{tabular} 
\caption{As \protect \ref{tab:vector} but for the $A_1^{-+}$ channel \label{tab:ps}}
     \end{table*}

\begin{figure*}
 \includegraphics[width=8cm,bb=0 0 865 586]{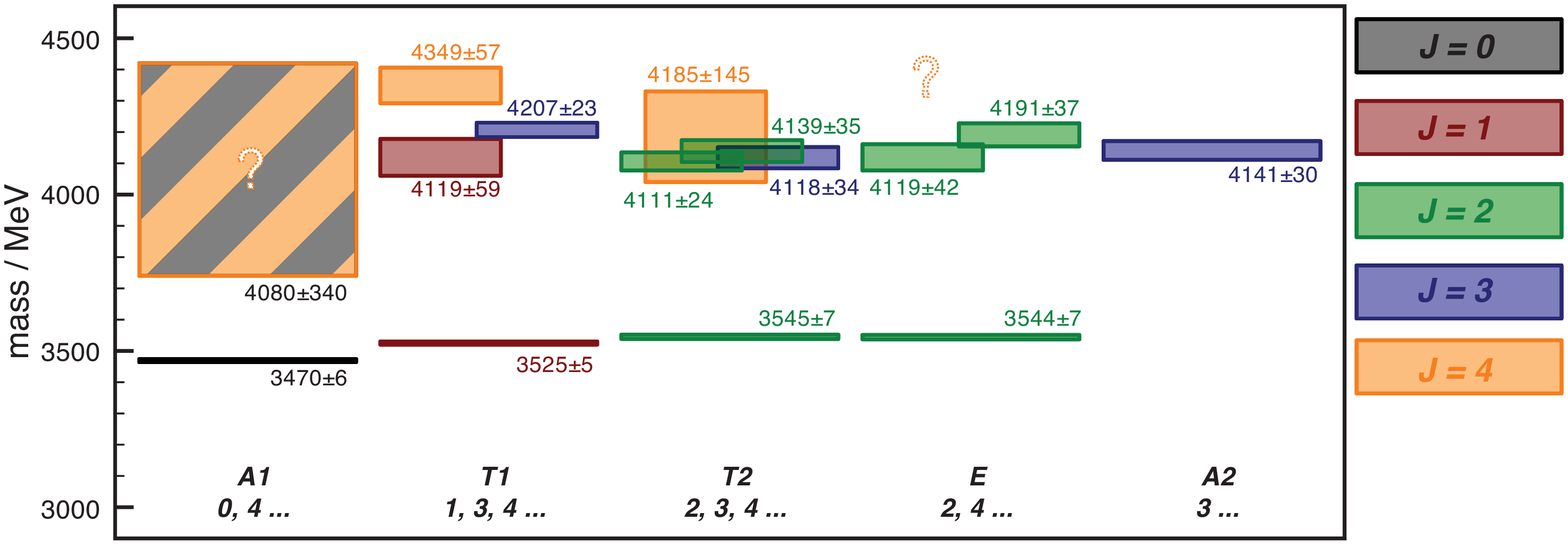}
\includegraphics[width=8cm,bb=0 0 729 295]{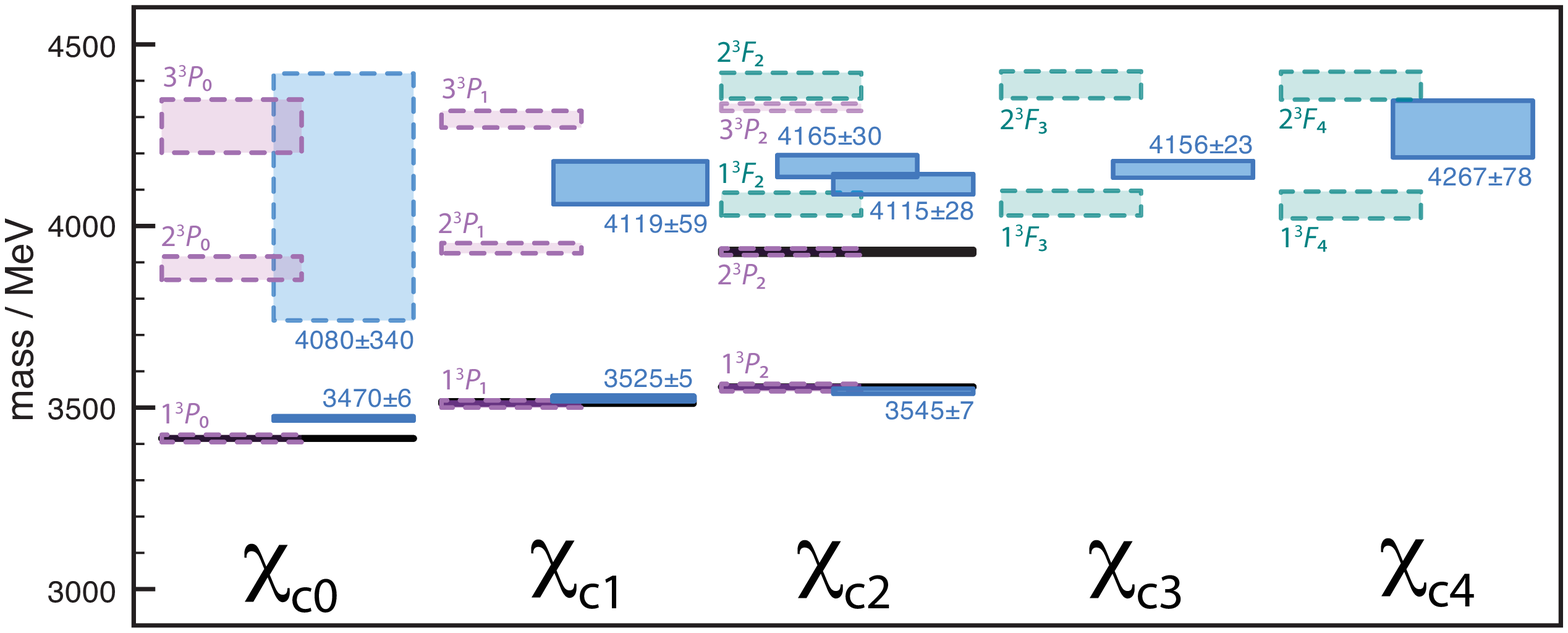}
\caption{$PC=++$ spectrum by lattice irrep and by continuum assignment. \label{fig:plusplus}}
\end{figure*}

Consideration of the channels $(A_1, T_1, T_2, E)^{++}$  gives us an insight into the common properties of spin-orbit split multiplets, e.g. $^3P_J$. The spectrum presented here differs slightly from that published in \cite{Dudek:2007wv} owing to the use of a somewhat larger operator basis causing an anticipated increase in the density of states produced through lattice discretisation artifacts. There is a clear ground state $1^3P_{J=0,1,2}$ multiplet split by small spin-orbit forces and above that there is a dense spectrum of states whose overlap extractions are somewhat noisy; for reasons of space we do not show the numbers here, displaying only the mass spectrum in Figure \ref{fig:plusplus}. A plausible description would be that:
\begin{itemize}
 \item The $A_1$ channel may house the $2^3P_0$ state at $4080(340)$, although the noisy overlaps of this state are consistent with being a $4^{++}$ artifact;
\item The $T_1$ channel houses the $2^3P_1$ state at $4119(59)$ and lattice artifact states at $4207(23), 4349(57)$ that might be the $1^3F_3(3^{++})$ and $1^3F_4(4^{++})$ states of the quark model;
\item The $T_2^{++}$ channel has four closely spaced levels around $4.1$ GeV that have a possible interpretation as the $2^3P_2, 1^3F_2, 1^3F_3, 1^3F_4$ states, only the last of which is a lattice artifact with the set of operators used;
\item The $E^{++}$ channel excited states near $4.1$ GeV have large (but noisy) overlaps suggesting that they are not lattice artifacts and are likely to be the $2^3P_2$ and $1^3F_2$ states. The reason for the non-appearance of a lattice artifact $4^{++}$, seen in $T_1$ and $T_2$, is not known.
\end{itemize}
We note that the overlaps onto the gluonic operator $b_1\times \mathbb{B}$ are uniform across all the states indicating within the quark model interpretation that there are no dominantly hybrid states in this mass region - if our interpretation is correct this may favour the Coulomb-gauge model of heavy-quark hybrids over the simplest flux-tube model. In the Coulomb-gauge model \cite{Guo:2008yz} the lightest $(0,1,2)^{++}$ hybrid states are heavier than the $1^{--}$ hybrid state and are quark spin-triplets, while in the simplest flux-tube model there is a spin-singlet $1^{++}$ hybrid state degenerate with the $1^{--}$ hybrid state, which we have proposed is located at a mass near $4.4$ GeV, and since the $b_1\times \mathbb{B}$ operator has a quark spin-singlet component it seems reasonable to anticipate overlap with this state. 

Finally we will consider applying our methods to channels that in continuum house $J^{PC}$ exotic states not present in the simple $\bar{c}c$ quark model. The most straightforward choice is $A_1^{+-}$ which only houses even spins and hence is always exotic with this $PC$. The simplest operator transforming in this way is $\bar{\psi}\gamma^0 \psi$ - we find that unsmeared correlators containing this operator are consistent with zero, which matches with the fact that this operator contains no gluonic field to generate a hybrid state. By smearing this operator using a function of the gauge-covariant laplacian we introduce the gluonic field and have the possibility of overlap with a hybrid state. Using this operator and smeared and unsmeared $a_1 \times \mathbb{B}$ operators we extract a state at $4465(65)$ MeV which has clear overlaps on to all three operators. The first excited state is much higher in mass, at $5570(270)$ MeV.

A less trivial channel is $T_1^{-+}$ which houses exotic $1^{-+}, 3^{-+}$ but also non-exotic $4^{-+}$ which is realised in the quark model as $^1G_4$. In \cite{Dudek:2007wv} it proved to be not possible to decisively state whether the ground state in this channel was indeed the exotic $1^{-+}$ state or a $4^{-+}$ nonexotic, here we will consider this again using our model-dependent overlap comparison. In Table \ref{tab:T1-+} we present the extracted overlaps using an eight dimensional basis of operators.
For the ground state we see rather large overlaps with the quark-model forbidden operators $\rho \times \mathbb{B}, \rho_2 \times \mathbb{B}$ suggesting that it may well be the exotic hybrid $1^{-+}$ state. The first excited state has overlaps consistent with zero for all operators except the smeared $b_1\times \nabla$ and possibly the $a_{0(2)}\times \nabla$. This may well be a signal for a lattice artifact $4^{-+}$ state, whose spin-singlet ($^1G_4$) nature could appear through overlap with $b_1 \times \nabla$ which has a dominant spin-singlet piece in its non-relativistic reduction. 

\begin{table*}
\begin{tabular}{c|cccc}
 operator & $0^{\mathrm{th}}[4305(40)]$ &  $1^{\mathrm{st}}[4645(86)]$ &  $2^{\mathrm{nd}}[4689(138)]$  &  $3^{\mathrm{rd}}[5580(160)]$ \\
\hline \hline
$a_{0(2)} \times \nabla_{T1}^{(sm)}\;(10^{-3})$ 	&  2.5(2)	& 2.0(6) 	& 2.0(7)	& 0.8(4) \\
$b_1 \times \nabla_{T1}^{(sm)}\;(10^{-3})$ 		& 2.2(2)	& 2.9(4)	& 1.7(8)	& 1.4(7)\\
$\rho \times \mathbb{B}_{T1}^{(sm)}\;(10^{-3})$ 			&  \textbf{2.88(5)}	& \textit{0.2(2)}	& 0.8(5)	& 0(0.3) \\
$\rho_2 \times \mathbb{B}_{T1}^{(sm)}\;(10^{-3})$ 			&  \textbf{2.84(5)}	& \textit{0.0(2)}	& 0.8(5)	&0(0.3) \\

$a_{0(2)} \times \nabla_{T1}\;(10^{-3})$ 			&  1.8(1)	& \textit{0.2(3)}	& 1.2(5)	& 1.5(7) \\
$b_1 \times \nabla_{T1}\;(10^{-3})$ 				& 1.7(1)	&\textit{0.2(4)}	& 1.0(5)	& 0.8(11)\\
$\rho \times \mathbb{B}_{T1}\;(10^{-3})$ 					&  3.1(2)	& \textit{0.4(7)}	& 2.6(9)	& 3.6(20) \\
$\rho_2 \times \mathbb{B}_{T1}\;(10^{-3})$ 				&  3.0(2)	& \textit{0.2(6)}	& 2.5(9)	& 3.5(20) \\
\hline
assignment	& $1^{-+}$ hyb ? & $4^{-+}(^1G_4)$ ? & ? & ? 
\end{tabular} 
\caption{As \protect \ref{tab:vector} but for the $T_1^{-+}$ channel.. \label{tab:T1-+}}
     \end{table*} 

$2^{+-}$ exotics appear in $(T_2,E)^{+-}$. The $T_2$ receives contributions also from non-exotic $3^{+-}$ and indeed the ground state in that channel is identified as such. The first excited state in $T_2^{+-}$ matches with the ground state in $E^{+-}$ and in both cases large overlaps with the $a_1 \times \mathbb{B}$ operator are seen strongly suggesting that this is the $2^{+-}$ exotic at a mass of $4620(60)$.

\section{Finite momentum}
The non-relativistic quark-model states as constructed do not transform covariantly under boosts, but do transform properly under rotations in three-dimensions. This can lead to overlap on to more states than are allowed by Lorentz symmetry. For example, consider the operator $\bar{\psi} \gamma^\mu \partial^\nu \psi$ - insisting upon Poincar\'{e} invariance one has only the following overlaps
\begin{eqnarray}
	\langle 0| \bar{\psi} \gamma^\mu \partial^\nu \psi | 0^{++}(\vec{p},r) \rangle &=& Z g^{\mu\nu} + Z' p^\mu p^\nu \nonumber \\
	\langle 0| \bar{\psi} \gamma^\mu \partial^\nu \psi  | 1^{++}(\vec{p},r) \rangle &=& Z \epsilon^{\mu\nu \rho \sigma} p_\rho \in_\sigma\hspace{-1mm}(\vec{p},r)  \nonumber \\
	\langle 0| \bar{\psi} \gamma^\mu \partial^\nu \psi  | 2^{++}(\vec{p},r) \rangle &=& Z \in^{\mu\nu}\hspace{-1mm}(\vec{p},r), \nonumber 
\end{eqnarray}
so that if, as we do in the lattice calculation, one considers only the spatial derivatives, one has overlaps
\begin{eqnarray}
	\langle 0| \bar{\psi} \gamma^i \partial^j \psi | 0^{++}(\vec{p},r) \rangle &=& Z \delta^{ij} + Z' p^i p^j  \nonumber  \\
	\langle 0| \bar{\psi} \gamma^i \partial^j \psi  | 1^{++}(\vec{p},r) \rangle &=& Z \epsilon^{ijk} \left(p^k \in^0\hspace{-1mm}(\vec{p},r) - E\in^k\hspace{-1mm}(\vec{p},r)  \right)  \nonumber  \\
	\langle 0| \bar{\psi} \gamma^i \partial^j \psi  | 2^{++}(\vec{p},r) \rangle &=& Z \in^{ij}\hspace{-1mm}(\vec{p},r). \nonumber 
\end{eqnarray}
But note that this is not the most general set allowed by three-dimensional rotations, parity and charge conjugation, giving up on boost invariance we also are allowed overlaps
\begin{eqnarray}
	\langle 0| \bar{\psi} \gamma^i \partial^j \psi | 0^{-+}(\vec{p},r) \rangle &=& Z \epsilon^{ijk} p_k  \nonumber  \\
	\langle 0| \bar{\psi} \gamma^i \partial^j \psi  | 2^{-+}(\vec{p},r) \rangle &=& Z \epsilon^{ikl} \in^{jk}\hspace{-1mm}(\vec{p},r) p^l . \nonumber 
\end{eqnarray}

Within the quark model state construction, we can explicitly compute these overlaps at finite momentum finding
\begin{eqnarray}
 	Z(^3P_0) &=& \sqrt{2 E_{\vec{p}}} \sqrt{\pi} \int \tfrac{q^2dq}{(2\pi)^3} \tfrac{1}{\sqrt{2}} \tfrac{q}{3 m_q^2} \varphi(q)  \nonumber  \\
	Z'(^3P_0) &=& \sqrt{2 E_{\vec{p}}} \sqrt{\pi} \int \tfrac{q^2dq}{(2\pi)^3} \tfrac{8}{3} q\left( 1-  \tfrac{q^2}{4 m_q^2} - \tfrac{p^2}{16m_q^2}     \right) \varphi(q)  \nonumber  \\
	Z(^3P_1) &=&  \sqrt{2 E_{\vec{p}}} \sqrt{\pi} \tfrac{M}{E_{\vec{p}}^2}   \int \tfrac{q^2dq}{(2\pi)^3} \tfrac{4}{\sqrt{3}} q\left( 1+  \tfrac{q^2}{4 m_q^2} + \tfrac{p^2}{16m_q^2}     \right) \varphi(q)   \nonumber  \\
	Z(^3P_2) &=& \sqrt{2 E_{\vec{p}}} \sqrt{\pi}  \int \tfrac{q^2dq}{(2\pi)^3} \tfrac{8}{\sqrt{6}} q \left( 1+  \tfrac{q^2}{20 m_q^2} - \tfrac{p^2}{16m_q^2}     \right) \varphi(q)  \nonumber  \\
	Z(^1S_0) &=& \sqrt{2 E_{\vec{p}}} \sqrt{\pi}  \int \tfrac{q^2dq}{(2\pi)^3} \tfrac{\sqrt{2}}{3} \tfrac{q^2}{m_q^2} \varphi(q)  \nonumber  \\
	Z(^1D_2) &=& \sqrt{2 E_{\vec{p}}} \sqrt{\pi}  \int \tfrac{q^2dq}{(2\pi)^3} \tfrac{2}{\sqrt{15}} \tfrac{q^2}{m_q^2} \varphi(q), \nonumber 
\end{eqnarray}
so that, as anticipated, as well as the overlaps allowed by Lorentz symmetry there are also disallowed overlaps. This is an inherent weakness of the non-relativistic model that can only be remedied by constructing a fully Poincar\'{e} covariant bound-state scheme which poses a significant challenge to modellers.

\section{Summary}

We have presented a simple framework for comparison of lattice QCD spectroscopy and the quark model. It relies upon a non-relativistic reduction and as such is suitable in the heavy-quark sector. We have compared with recent lattice QCD data and presented a model-dependent description of the data which agrees in structure with the predictions of the Cornell potential quark model, but goes beyond that model in providing predictions for exotics and crypto-exotic hybrid mesons. The particular lattice data used is dominated by sources of systematic error, notably we suspect that the small volume used is ``squeezing'' the wavefunctions of higher excited states. An alternative use of this method, when applied to more realistic lattice QCD data, would be to allow a quark model to be ``tuned'' to QCD, through selection of interactions and parameters.

Of particular interest is the extension beyond quark-model states, where the gluonic field plays a manifest role. If the assignments of hybrid nature made in this analysis are correct we have the following (incomplete) hybrid spectrum:
\begin{itemize}
	\item a non-exotic pseudoscalar state ($0^{-+}$)  at $4280(60)$ MeV which may have a degree of mixing with a nearby conventional $\bar{c}c$ state;
	\item an exotic $1^{-+}$ state at $4305(50)$ MeV;
	\item a non-exotic vector state ($1^{--}$) around $4400(60)$ MeV where mixing with conventional states is not apparent;	
	\item an exotic $0^{+-}$ state at $4465(65)$ MeV;
	\item an exotic $2^{+-}$ state at $4620(60)$ MeV;
	\item no non-exotic hybrids in $(0,1,2)^{++}$ channels below about $4.5$ GeV 	
\end{itemize}
An interesting extension to the work done in this paper might be to apply a similar technique using a model with explicit gluonic degrees-of-freedom such as the flux-tube model or the Coulomb-gauge model.

It is not clear if this method will have utility for lighter quarks, where the quasiparticle quark-like degrees-of-freedom in the quark model (``constituent quarks'') are not the same quark degrees-of-freedom that appear in the QCD Lagrangian (``current quarks'').

\begin{acknowledgments}
Notice: Authored by Jefferson Science Associates, LLC under U.S. DOE
Contract No. DE-AC05-06OR23177. The U.S. Government retains a
non-exclusive, paid-up, irrevocable, world-wide license to publish or
reproduce this manuscript for U.S. Government purposes. 
Computations were performed on clusters at Jefferson Laboratory as part of the USQCD
collaboration.
E.R. recognises support from the JSA Research Internship for Foreign Undergraduates fund.
\end{acknowledgments}

\bibliography{charm_spectrum} 

\end{document}